\documentstyle[aps,graphicx,floats,twocolumn]{revtex}
\begin{document}
\draft
\wideabs
{

\title{A low-dimensional spin $S = 1/2$ system at the quantum critical
limit: Na$_{2}$V$_{3}$O$_{7}$}

\author{J. L. Gavilano$^{1}$,  D. Rau $^{1}$,S. Mushkolaj$^{1}$, H. R. Ott$^{1}$,
P. Millet$^{2}$, and F. Mila$^{3}$}

\address{
$^{1}$ Laboratorium f\"{u}r Festk\"{o}rperphysik,
ETH-H\"{o}nggerberg, CH-8093~Z\"{u}rich, Switzerland \\
$^{2}$ Centre d'Elaboration des Mat$\acute{e}$riaux et d'Etudes
Structurales, 29, rue J. Marvig, 31055 Toulouse Cedex, France \\
$^{3}$ Institut de Physique Th\'{e}orique, Universit\'{e} de
Lausanne, CH - 1015 Lausanne, Switzerland  }

\maketitle              
\begin{abstract}
We report the results of measurements of the dc-susceptibility and the
$^{23}$Na-NMR response of Na$_{2}$V$_{3}$O$_{7}$, a recently
synthesized, non metallic low dimensional spin system.  Our results
indicate that upon reducing the temperature to below 100 K, the
V$^{4+}$ moments are gradually quenched, leaving only one moment out
of 9 active. The NMR data reveal a phase transition at very low
temperatures.  With decreasing applied field $H$, the critical
temperature shifts towards $T = 0$ K, suggesting that
Na$_{2}$V$_{3}$O$_{7}$ may be regarded as an insulator reaching a
quantum critical point at $H = 0$.
\end{abstract}

\vfill
}

Low dimensional spin systems have recently been the subject of
interest in a large number of research projects.  The interest in
these systems is manifest in many theoretical studies and much
progress has been made in the synthesis of new compounds, which may be
considered to be physical realizations of systems with one- and
two-dimensional arrays of spins in the form of spin chains and
ladders\cite{Dagotto96}.  Prominent examples include Cuprates and
ternary Vanadium oxides\cite{Ueda98,Millet98}.  The physical
properties of ladder compounds are largely determined by the number of
legs of the individual ladders.  Less clear is the influence of the
topology of a system on its physical properties.  It has been
suggested\cite{Schulz96,Kawano97} that the topology, specifically the
periodic boundary conditions in the rung direction, is essential for
the low temperature properties of some ladder compounds.

In this respect, new opportunities emerged from recognizing that
variations in the composition of some two-dimensional layered
compounds result in microscopic tubular
structures\cite{Tenne92,Frey98,Seifert2001}.  In 1999 Millet and
coworkers\cite{Millet99} reported the synthesis of
Na$_{2}$V$_{3}$O$_{7}$, a material whose structure may be considered
as composed by nanotubes oriented along the c-axis of the crystal
lattice (space group P31c).  The individual nanotubes, with an inner
diameter of approximately 5 $\AA$, are formed by VO$_{5}$ square
pyramids, joined at their edges and corners.  The Na ions, occupying
different sites inside and around individual nanotubes, are bonded to
the O ions of up to three nanotubes.  From our experimental data we
infer that, upon reducing the temperature from 100 to 10 K, the
effective moment of the V$^{4+}$ ions in Na$_{2}$V$_{3}$O$_{7}$ is
reduced by one order of magnitude.  Below 20 K our data reveal that
for $H = 0$, this material is close to a phase transition at $ T=0 $
K.

\begin{figure}[ht]
\vspace{2ex plus 1ex minus 0.5ex}
\begin{center}\leavevmode
\includegraphics[width=0.8 \linewidth]{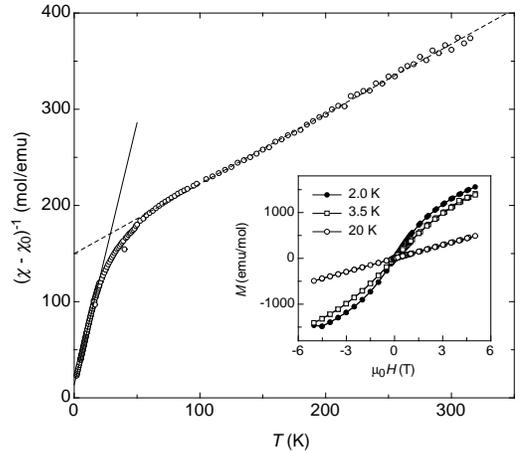}
\caption{ Temperature dependence of the inverse magnetic susceptibility
measured at 0.1 T. Below 20 K, also  data measured in an
applied field of 1 T are included (the two data sets cannot be distinguished).  The
broken and solid line represent the best fits to a Curie-Weiss type
behavior at high and low temperature, respectively.  Inset:
Magnetization $M(H)$ curves at three different temperatures. } \protect
\label{Figure1}
\end{center}
\end{figure}

Our Na$_{2}$V$_{3}$O$_{7}$ samples consisted of ensembles of randomly
oriented small crystals, with typical dimensions of 10$\mu
$m$\times$10$\mu $m$\times$1mm.  The material was grown under vacuum
from melts with a starting composition of
Na$_{1.9}$V$_{3}$O$_{7}$\cite{Millet99}.  The susceptibility was
measured using a commercial SQUID magnetometer and the NMR
measurements employed standard spin-echo techniques.  In Fig.  1 we
display the temperature dependence of the molar magnetic
susceptibility $\chi(T)$ of Na$_{2}$V$_{3}$O$_{7}$, at temperatures
between 1.9 and 315 K. Curie-Weiss type features of $\chi(T)$ of the
form
\begin{equation}
  \chi(T)  = \chi_{dia} + \frac{C}{(T-\Theta)} \; .
\label{eq:1}
\end{equation}
 are discernible in two different temperature regimes.  The background
 $\chi_{dia}$ was extracted from the high temperature data.  Above 100
 K, the best fit to the data, represented by the broken line in Fig. 
 1, yields $C \approx 1.4$ emu$\cdot$K/mol and $\Theta = -200 \pm 30 $
 K. These values imply an effective magnetic moment of 1.9 $\mu_{B}$
 per V ion, close to the expected value for V ions in their
 tetravalent configuration, and substantial antiferromagnetic
 interactions between them.  Below approximately 100 K, $\chi(T)$
 changes gradually and at much lower temperatures, between 20 K and
 1.9 K, the straight solid line in Fig.  1 implies that, with respect
 to the high-temperature values, $C$ and $\Theta$ are reduced by
 approximately one and two orders of magnitude, respectively.  These
 results confirm those obtained previously in higher external
 fields\cite{Gavilano2002}.  Considering these data and magnetization
 curves $M(H)$ displayed in the inset of Fig.  1,
 Na$_{2}$V$_{3}$O$_{7}$ exhibits neither a phase transition to a
 magnetically ordered state nor spin freezing phenomena at
 temperatures above 2 K. The reduction of the effective magnetic
 moment is most likely due to a gradual process of moment compensation
 via the formation of singlet spin configurations with most but not
 all of the ions taking part in the process.  This may be the result
 of antiferromagnetic interactions and geometrical frustration.  It is
 tempting to conjecture the compensation of 8 out of the 9 V spins
 along the ``circumference'' of the individual nanotubes of
 Na$_{2}$V$_{3}$O$_{7}$.  This is natural if the dominant interactions
 are between edge-sharing VO$_5$ pyramids, but more difficult to
 understand if the dominant interactions are between corner-sharing
 pyramids, as suggested in Ref.\cite{Whangbo00}.

\begin{figure}[ht]
\vspace{2ex plus 1ex minus 0.5ex}
\begin{center}\leavevmode
\includegraphics[width=0.8 \linewidth]{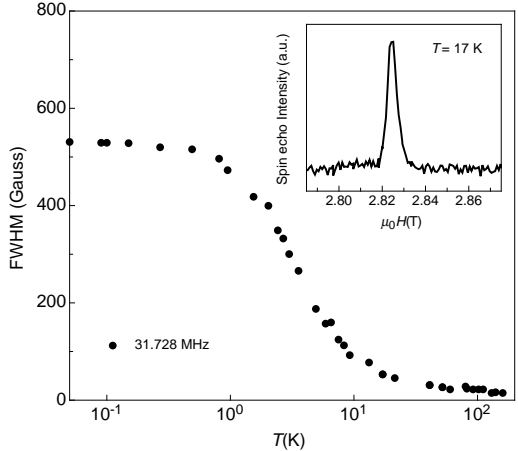}
 \caption{ $^{23}$Na-NMR line width ($FWHM$) as a function of temperature
 measured at 31.728 MHz.  A rather large (and
 field dependent) increase of the linewidth is observed below 30 K.
 Inset:  $^{23}$Na-NMR spectrum measured at 17 K. }
 \protect
\label{Figure2}
\end{center}
\end{figure}

The $^{23}$Na-NMR response of Na$_{2}$V$_{3}$O$_{7}$ was measured in
different external magnetic fields.  An example of the $^{23}$Na-NMR
spectra is displayed in the inset of Fig.  2, measured at a fixed
Larmor frequency of 31.728 MHz.  No quadrupolar wings appear in our
data because their intensity is distributed over a broad frequency
range\cite{FieldGradients}.  In the inset of Fig.  2, only the
unresolved signals of the Zeeman central transitions $(+1/2
\leftrightarrow -1/2)$ from the 4 Na sites are observed near 2.825 T,
suggesting rather modest differences, if any, in the hyperfine fields
at the different Na sites.  This is is consistent with a direct
dipolar coupling between the V$^{4+}$ moments and the Na nuclei (see
below).

 In the main frame of Fig.  2 we display the temperature evolution of
 the $^{23}$Na-NMR linewidth ($FWHM$), measured at a fixed frequency
 of 31.728 MHz.  Between 50 to 150 K the NMR linewidth increases
 gradually with decreasing temperature by approximately a factor of 2.
 A more dramatic change occurs at lower temperatures where an increase
 of more than one order of magnitude is observed as the temperature is
 reduced to below 1 K. This type of behavior usually reflects
 magnetic-ordering or spin-freezing phenomena.  In our case, the
 temperature-independent linewidth at low temperatures increases with
 increasing applied field (data not shown), not consistent with either
 a simple ferro- or antiferromagnetic order.

\begin{figure}[ht]
\vspace{2ex plus 1ex minus 0.5ex}
\begin{center}\leavevmode
\includegraphics[width=0.8 \linewidth]{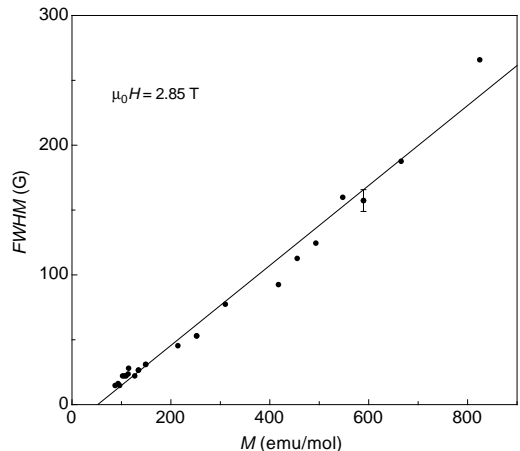}
\caption{ $FWHM$ as a function of the molar magnetization $M$, at the
chosen field and at different temperatures.  The linear relation
between $FWHM$ and $M$ indicates that the increase of the linewidth is
of magnetic origin.}
\protect
\label{Figure3}
\end{center}
\end{figure}

In Fig.  3 we display the full $^{23}$Na NMR linewidth at half maximum
($FWHM$) as a function of the molar magnetization $M$ for temperatures
above 3.5 K. Note the linear relation $ FWHM = (-16 + 0.308\cdot M)$
Gauss.  The NMR linewidth is caused by a distribution of internal
static fields at the four inequivalent Na sites in randomly oriented
grains.  These fields arise from the polarization of the V moments in
an external field and therefore, we assume $FWHM \approx \Delta
A_{hf}(M/\mu_{B}N_{A})$ where $M$ is the molar magnetization, $\Delta
A_{hf}$ is the width of the distribution of hyperfine field couplings,
$\mu_{B}$ and $N_{A}$ are the Bohr magneton and Avogadro's number,
respectively.  From the slope of the straight solid line in Fig.  3 we
infer that $\Delta A_{hf} = 1.7$ kOe per $\mu_{B}$ of V moment.

In order to gain a numerical estimate we have calculated the dipolar
fields at the 4 Na sites generated by equally polarized V moments of 1
$\mu_{B}$, with the polarization along several directions.  The
results\cite{DipolarFields} suggest that for a randomly oriented
powdered sample, the total width of the $^{23}$Na-NMR signal ought be
of the order of 2.5 kOe.  The experimental value $\Delta A_{hf} = 1.7$
kOe per $\mu_{B}$ of V moment is thus consistent with a direct dipolar
hyperfine coupling between the V moments and the Na nuclei.  The
quoted result still applies, if the formation of singlet states occurs
for most of the V moments in clusters of ions, as long as the
uncompensated spins are mobile in these clusters.  The characteristic
frequencies of this motion need to exceed the NMR Larmor frequencies. 
If, however, after the singlet formation the arrangements of spins are
static, the dipolar fields are expected to vary depending on the
details of the arrangement.  Preliminary estimates indicate that
although the fields at individual Na sites change substantially, the
spread of their values is of the same order of magnitude as mentioned
above.  This analysis suggests that the S=1/2 ground-state of the
9-spin ring, with one S=1/2 spin delocalized into a sea of four
singlets, is the relevant effective degree of freedom at low energy.

\begin{figure}[ht]
\vspace{2ex plus 1ex minus 0.5ex}
\begin{center}\leavevmode
\includegraphics[width=0.8 \linewidth]{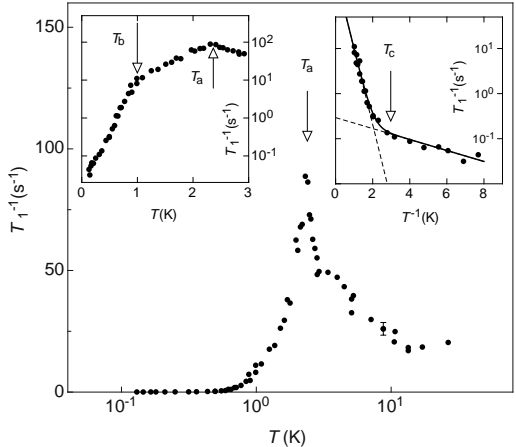}
\caption{ Main frame: $T_{1}^{-1}(T)$ measured at 76.5 MHz below 40 K.
A peak is observed at $T_{a}= 2.35$ K. Inset left: A slope change in
$T_{1}^{-1}(T)$ is revealed at $T_{b}= 1$ K. Inset right: A third
anomaly in $T_{1}^{-1}(T)$ appears at $T_{c}= 0.43$ K. The solid line
represents the sum of two exponential functions.  } \protect
\label{Figure4}
\end{center}
\end{figure}

We now turn to the results of measurements of the spin-lattice
relaxation rate $T_{1}^{-1}$ as a function of temperature in various
external fields.  The values of $T_{1}^{-1}$ were extracted from fits
to nuclear magnetization recovery curves $m(t)$, obtained as follows.
First the populations of the central $^{23}$Na nuclear Zeeman levels
($I = +1/2, -1/2$) were equalized by applying a long comb of
rf-pulses.  After a variable delay $t$ a $\pi/2 - \pi$ spin-echo
sequence was applied and the corresponding spin-echo intensity,
$m(t)$, was recorded.  In the main frame of Fig.  4 we display
$T_{1}^{-1}(T)$ measured in a magnetic field of 68.13 kOe below 40 K.
The complete data set, covering temperatures up to 300 K gives no
indication of a gap in the spin excitation spectrum at temperatures
above the liquid He temperature range.  Besides the peak at $T_{a}=
2.3 $K (see Fig.  4), two additional significant anomalies in
$T_{1}^{-1}(T)$ appear at $T_{b} = 1$ K (left inset) and at $T_{c} =
0.43$ K (right inset).  Below $T_{b}$, $T_{1}^{-1}(T) \propto
exp(-\Delta/k_{B}T)$ with values $\Delta/k_{B}$ of 3.8 and 0.28 K above
and below $T_{c}$, respectively.  The evolution of these features with
magnetic field below 10 K is depicted in Fig.  5.  This suggests a
small gap in the spin excitation spectrum at these temperatures.
Since we are dealing with an $S = 1/2$ spin system any anisotropy
indicates some coupling to other degrees of freedom.  The anomalies in
$T_{1}^{-1}(T)$ are indicated by arrows at $T_{a}$, and by the two
dotted lines for $T_{b}$ and $T_{c}$.  Substantial field-induced
changes in $T_{1}^{-1}(T)$ are observed at low temperatures,
confirming that the state below $T_{a}$ involves spin degrees of
freedom.  As is shown in the inset of Fig.  5, the transition
temperature $T_{a}$, the onset of the spin gap at $T_{b}$ and the
manifestation of the anisotropy by the feature at $T_{c}$ are all
smoothly shifting to zero upon reducing the external field to zero.
This indicates that Na$_{2}$V$_{3}$O$_{7}$, an insulator, may be
considered as being close to or at a quantum critical point at $H =
0$.

Features of $T_{1}^{-1}(T)$ as those shown in the main frame of Fig.
4 are usually associated with a cooperative phase transition which, in
magnetic systems, involves spin (and in some cases lattice) degrees of
freedom.  At this point we assume that a cooperative phase transition
occurs at $T_{a}$.  From $\chi(T)$ at temperatures
between 2 and 20 K, the related paramagnetic Curie temperature for
Na$_{2}$V$_{3}$O$_{7}$ is $\Theta = -2.5$ K. Hence it is tempting to
conclude that a transition to an antiferromagnetic state occurs at
$T_{a}$.  However, as stated above, the behaviour of the line
width is not consistent with antiferromagnetic ordering.
Besides, the temperature dependence of $T_{1}^{-1}$ below
$T_{a}$ is not consistent with expectations for a common magnetically
ordered system.  In particular, the change of slope in $T_{1}^{-1}(T)$
at $T_{b}$ suggests that the formation of a gap in the spin excitation
spectrum of the ordered state does not occur at the transition,
$T_{a}$, but only at the lower temperature $T_{b}$.  This clearly does
not match with $T_{1}^{-1}(T)$ of ordinary varieties of ferro- or
antiferromagnetically ordered systems.

\begin{figure}[ht]
\vspace{2ex plus 1ex minus 0.5ex}
\begin{center}\leavevmode
\includegraphics[width=0.8 \linewidth]{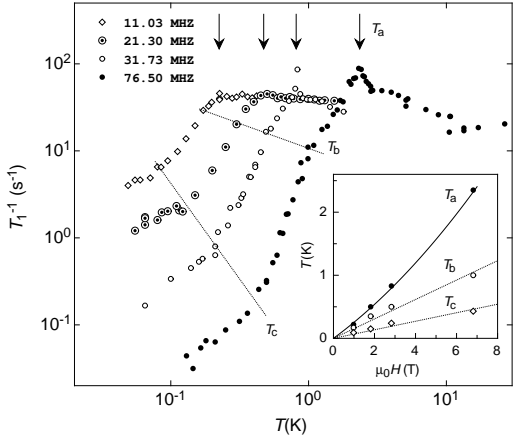}
 \caption{ $T_{1}^{-1}(T)$ for external fields of 9.8, 18.1, 28.25 and
 68.13 kOe, corresponding to Larmor frequencies of 11.03, 21.30 31.728
 and 76.5 MHz, respectively.  Anomalies in $T_{1}^{-1}(T)$ are
 emphasized by arrows at $T_{a}$ and by two broken lines indicating $T_{b}$
 and $T_{c}$.  Inset: Magnetic field dependence of $T_{a}$, $T_{b}$
 and $T_{c}$.  The different lines are to guide the eye.  } \protect
\label{Figure5}
\end{center}
\end{figure}

These and previous arguments lead to the conclusion that in the
observed phase transition, degrees of freedom other than those of the
spin system are involved.  The coupling of spins to lattice
distortions often leads to complications.  A recent example is
Cu$_2$(C$_5$H$_{12}$N$_2$)$_2$Cl$_4$, a spin ladder
system\cite{Chaboussant97} where magnetoelastic effects are believed to
influence a rather rich $[H,T]$ phase diagram at low
temperatures\cite{Stone02,Mayaffre00}.  In this case, two quantum critical
points at two different external magnetic fields $H_{c1}$ and $H_{c2}$
have been identified\cite{Mayaffre00}.  Since our data suggest a
quantum critical point at or very close to $H = 0$, any intrinsic gap
in the V spin excitation spectrum of the paramagnetic phase ought to be tiny.  This is
compatible with the topology of the spin arrangement in
Na$_{2}$V$_{3}$O$_{7}$ and the observed reduction to one effective
spin 1/2 per turn around the tubes at low temperatures. This type of
spin tubes is generically expected to have a much smaller gap
than even-leg ladders.
This is particularly clear in the limit where the rung coupling 
$J_\perp$ is much larger than the leg coupling $J_\parallel$, 
presumably the relevant one for both systems: In two-leg
ladders, the gap opens on each rung and scales with
$J_\perp$\cite{Dagotto96}, while in odd-leg ladders with
 periodic boundary conditions,
the gap opens due to a dimerization along the tube and scales
with $J_\parallel$\cite{Schulz96,Kawano97}.

In summary, from the results of our dc-susceptibility and
$^{23}$Na-NMR measurements on Na$_{2}$V$_{3}$O$_{7}$, we conclude that
at high temperatures, all the V ions are in a tetravalent state and
all related moments contribute to the magnetic susceptibility.  This
situation changes gradually below 100 K. Below 20 K a different
paramagnetic state is established with only 1 out of 9 of the V
magnetic moments contributing to the susceptibility.  At much lower
temperatures and in the presence of modest external magnetic fields, a
phase transition occurs at a field-dependent transition temperature
$T_{a}$.  In the ordered phase, an anisotropic gap $\Delta$ forms at
temperatures distinctly below $T_{a}$.  All energy scales
characterizing this ordered state, $i.e.$, $T_{a}$, $T_{b}$ and
$T_{c}$, shift towards $T= 0 $ K with decreasing external field $H$. 
We therefore conclude that Na$_{2}$V$_{3}$O$_{7}$ may be considered as
being very close to or at a quantum critical point at $ H = 0$.  It is
conceivable that this quantum critical point is analogous to the
one that appears in gapped systems when the field is large enough to
close the gap.

This work was financially supported
by the Schweizerische Nationalfonds zur F\"{o}rderung der
Wissenschaftlichen Forschung.




\begin{thebibliography}{99}
\bibitem{DipolarFields} For the considered site configurations the extremal
values of the dipolar magnetic fields are 1630 and -770 Oe,
respectively.
\end{thebibliography}
\end{document}